\documentclass[twocolumn,showpacs,preprintnumbers,amsmath,amssymb
,superscriptaddress]{revtex4}

\usepackage{graphicx}
\usepackage{dcolumn}
\usepackage{bm}

\begin{document}

\title{Observation of an oscillatory correlation function 
of multimode two-photon	pairs}

\author{Hayato Goto}
\affiliation{Core Research for Evolutional Science and Technology(CREST), 
Japan Science and Technology Corporation (JST)} 
\affiliation{Department of Physics, Graduate School of Science, 
University of Tokyo, 
7-3-1 Hongo, Bunkyo, Tokyo, 113-0033, Japan}
\author{Yasuo Yanagihara}
\affiliation{Core Research for Evolutional Science and Technology(CREST), 
Japan Science and Technology Corporation (JST)} 
\affiliation{Department of Physics, Graduate School of Science, 
University of Tokyo, 
7-3-1 Hongo, Bunkyo, Tokyo, 113-0033, Japan}
\author{Haibo Wang}
\affiliation{Core Research for Evolutional Science and Technology(CREST), 
Japan Science and Technology Corporation (JST)} 
\author{Tomoyuki Horikiri}
\affiliation{Core Research for Evolutional Science and Technology(CREST), 
Japan Science and Technology Corporation (JST)} 
\affiliation{Department of Physics, Graduate School of Science, 
University of Tokyo, 
7-3-1 Hongo, Bunkyo, Tokyo, 113-0033, Japan}
\author{Takayoshi Kobayashi}
\affiliation{Core Research for Evolutional Science and Technology(CREST), 
Japan Science and Technology Corporation (JST)} 
\affiliation{Department of Physics, Graduate School of Science, 
University of Tokyo, 
7-3-1 Hongo, Bunkyo, Tokyo, 113-0033, Japan}

\date{\today}


\begin{abstract}
An oscillatory correlation function has been observed 
by the coincidence counting of 
multimode two-photon pairs produced with 
a degenerate optical parametric oscillator far below threshold.
The coherent superposition of the multimode two-photon pairs 
provides the oscillation in the intensity correlation function. 
The experimental data are well fitted to a theoretical curve. 
\end{abstract}

\maketitle

Parametric down-conversion(PDC) is one of the most important 
resources in quantum information because of the ability to generate 
correlated two-photon pairs. 
The correlation of two photons produced by PDC has been 
studied theoretically
and 
experimentally
\cite{Weinberg,MandelPDCtheoryA,MandelPDCtheoryL}. 
The correlated two-photon states have been utilized 
to realize quantum interference
\cite{Ghosh,Hong,Ou1989,Ou1990,OuFranson,Chiao,Zou}, 
quantum teleportation\cite{BouwmeeterTeleportation},  
ghost imaging\cite{GhostImaging}, and 
quantum lithography\cite{ShihLithography}. 
On the other hand, an optical parametric oscillator(OPO), which 
consists of a nonlinear crystal inside a cavity, 
has been investigated theoretically\cite{Milburn,Collett,GardinerOC} and 
experimentally\cite{KimbleSqueeze} as a squeezed-state source. 
Ou and Lu have recently used an OPO as a two-photon source
\cite{OuPRL,OuPRA}. 
Two photons produced with an OPO far below threshold 
have a narrow bandwidth limitted by that of the OPO cavity. 
The narrow-band two-photon state from an OPO 
is advantageous in experiments of interference because of the ability 
to provide high visibility\cite{Ou}. 
Ou and Lu have observed nonclassical photon statistics 
due to quantum interference between the narrow-band two-photon state 
and a coherent state\cite{OuStatistics}.

In Ref.\cite{OuPRA}, the multimode structure of the output from an 
OPO has also been discussed. 
The intensity correlation function of the multimode two-photon state
derived in Ref.\cite{OuPRA} 
is oscillatory. 
However, the coincidence rate of the multimode output 
reported\cite{OuPRA} is similar to 
that of the single-mode output 
because of the shorter round-trip time of the OPO cavity than 
the resolving time of detectors. 
In this Report, we report the observation of the oscillatory
correlation function of the multimode two-photon state 
from a relatively long OPO. 
The experimental data of the coincidence counting
are well fitted to a theoretical curve.

The output of a degenerate OPO far below threshold 
is composed of multimode two-photon pairs\cite{OuPRA}. 
The output operator of the OPO is given by the following equation
\cite{GardinerOC,OuPRA}:
\begin{widetext}
\begin{align}
a_{out}(\Omega ) 
= 
\sum_{m=-N}^N 
\left[
G_1(\Omega -m\Omega_F) a_{in} (\Omega )
+
g_1(\Omega -m\Omega_F) a^{\dagger}_{in} (-\Omega )
\right.
+
\left.
G_2(\Omega -m\Omega_F) b_{in} (\Omega )
+
g_2(\Omega -m\Omega_F) b^{\dagger}_{in} (-\Omega )
\right],
\label{OPOout}
\end{align}
\end{widetext}
with
\begin{align}
G_1({\Omega}) 
&=
\frac{\gamma_1 - \gamma_2 + 2i\Omega}{\gamma_1 + \gamma_2 - 2i\Omega}
,~
g_1({\Omega})
=
\frac{4\epsilon \gamma_1}{(\gamma_1 + \gamma_2 -2i\Omega )^2},
\nonumber
\\
G_2({\Omega}) 
&=
\frac{2\sqrt{\gamma_1 \gamma_2}}{\gamma_1 + \gamma_2 - 2i\Omega}
,~
g_2({\Omega})
=
\frac{4\epsilon \sqrt{\gamma_1 \gamma_2}}{(\gamma_1 + \gamma_2 -2i\Omega )^2}.
\label{coefficients}
\end{align}
Here $a_{in} (\Omega )$ and $b_{in}(\Omega )$ represent the vacuum mode 
entering the OPO cavity through an output coupler and 
the unwanted vacuum mode corresponding to other losses than 
the output coupler, respectively.
The frequency of these modes is
$\omega_0 + \Omega$ ($\omega_0$:the degenerate frequency of the OPO). 
$\gamma_1$ and $\gamma_2$ are the coupling constants for 
$a_{in}$ and $b_{in}$, respectively.
$\Omega_F$ is the free spectral range(FSR) of the OPO.
$2N+1$ is the number of the longitudinal modes in the output of the OPO.
$\epsilon$ is the single-pass parametric amplitude gain, 
which is proportional to the pump amplitude and the nonlinear coefficient. 
The intensity correlation function is defined as
\begin{align}
\Gamma (\tau )
=
\langle 
E^{(-)}(t) 
E^{(-)}(t+\tau) 
E^{(+)}(t+\tau) 
E^{(+)}(t) 
\rangle,
\label{correlationdefine}
\end{align}
with
\begin{align}
E^{(+)}(t) 
=
[E^{(-)}(t)]^{\dagger} 
=
\frac{1}{\sqrt{2\pi}}
\int
d\Omega a_{out} (\Omega ) e^{-i(\omega_0 + \Omega) t}.
\label{fielddefine}
\end{align}
From Eqs. (\ref{OPOout}), (\ref{coefficients}), (\ref{correlationdefine}), 
and (\ref{fielddefine}), we obtain 
\begin{align}
\Gamma (\tau )
&=
|\epsilon |^2 \left( \frac{F}{F_0} \right)^2 
\left\{
\left( 
\frac{2|\epsilon | (2N+1)}{\Omega_c}
\right)^2
\right.
\nonumber
\\
&+
\left.
e^{-\Omega_c |\tau | }
\frac{\sin^2 [(2N+1) \Omega_F \tau /2]}{\sin^2 (\Omega_F \tau /2)}
\left[ 
1
+
\left( 
\frac{2|\epsilon |}{\Omega_c}
\right)^2
\right]
\right\}.
\label{correlation}
\end{align}
Here $F$ and $F_0$ are the finesse of the OPO with and without loss, 
respectively; 
$\Omega_c$ is the bandwidth of the OPO. 
We retain
the terms of higher order than $|\epsilon |^2$ which were dropped 
in Ref.\cite{OuPRA} 
to explain our experimental results in further detail. 
The first term in Eq.(\ref{correlation}), 
which is independent of the delay time $\tau$, 
corresponds to the effect of two or more two-photon pairs. 
Each two-photon pair is uncorrelated with the other pairs. 
Therefore one photon of a pair and another one of the other pair can give
a coincidence count independent of the delay time. 
This first term represents the coincidence counts 
due to this process. 
From Eq.(\ref{correlation}), we find 
the time-domain comb-like structure of
the correlation function with the interval $\tau_F = 2\pi /\Omega_F$ of 
the peaks. $\tau_F$ is the round-trip time of the OPO cavity. 
In experiments, however, 
an observed correlation function is 
an average of Eq.(\ref{correlation}) 
over the resolving time, $T_R$, of detectors. 
Therefore, we cannot observe the comb-like correlaiton function 
if $T_R$ is longer than $\tau_F$\cite{OuPRA}. 
In our experiment, the inequality $T_R < \tau_F$ is satisfied and 
this makes possible 
to observe the oscillatory correlation function. 
We express the probability distribution of the timing jitter of detectors 
as $p(\tau )$. 
That is, $p(\tau )$ denotes the probability that a signal is 
output from a detector at time $\tau$ when a photon comes into 
the detector at time 0. The averaged intensity correlation function 
$\bar{\Gamma} (\tau )$
is given by
\begin{align}
\bar{\Gamma}(\tau ) 
&=
\int d\tau_1 d\tau_2 \Gamma (\tau_1 ) p(\tau_2 ) p(\tau_2 + \tau -\tau_1).
\label{correlationav1}
\end{align}
When $T_R \gg \tau_F/(2N+1)$, 
we can approximate the square of the ratio between the two sine functions in 
Eq.(\ref{correlation}) to a sum of delta functions in calculating 
$\bar{\Gamma} (\tau )$. 
Then Eq.(\ref{correlationav1}) becomes 
\begin{align}
\bar{\Gamma}(\tau ) 
&=
|\epsilon |^2 \left( \frac{F}{F_0} \right)^2 
\left[
\left( 
\frac{2|\epsilon | (2N+1)}{\Omega_c}
\right)^2
\right.
\nonumber
\\
&+
\left.
\tau_F (2N+1) 
e^{-\Omega_c|\tau | }
\sum_n 
\int d\tau' p(\tau' ) p(\tau' + \tau - n\tau_F)
\right], 
\label{correlationav2}
\end{align}
where we have dropped $(2|\epsilon |/\Omega_c)^2$ 
because it is much smaller than 1 in operating an OPO far below threshold. 
The coincidence measured in experiments will be proportional to 
$\bar{\Gamma} (\tau )$ given by Eq.(\ref{correlationav2}).

\begin{figure}[hbp]
	\includegraphics[width=8cm, height=5.5cm]{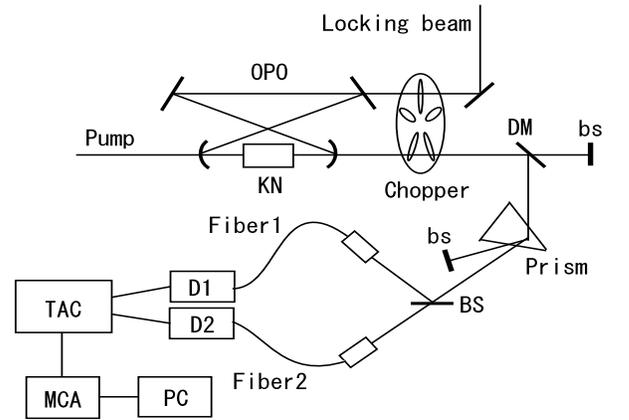}
	\caption{Schematic of the experimental setup. 
					KN:KNbO$_3$ crystal; DM:dichroic mirror; 
					BS:50/50 beamsplitter;
					bs:beam stop;
					D1 and D2:avalanche photodetectors; 
					TAC:time-to-amplitude converter; 
					MCA:multichannel analyzer.}
	\label{scheme}
\end{figure}

The schematic of the experiment for the observation of the 
oscillatory correlation function is shown in Fig.\ref{scheme}. 
A single-mode cw Ti:Sapphire laser of wavelength 860nm is used, 
which is pumped by the second harmonic of diode-pumped  YAG laser. 
The OPO cavity used in our experiment is a bow-tie ring cavity 
composed of two concave mirrors of curvature radius 50mm and 
two plane mirrors. 
The short and long path lengths bewteen the two concave mirrors are about 
60mm and 500mm, respectively. 
The output coupler of the OPO has about 10-\% and 85-\% transmittance 
at 860nm and 430nm, respectively. 
The other mirrors have high reflectance and transmittance at 860nm 
and 430nm, respectively. 
The nonlinear crystal in the OPO is a 10-mm-long a-cut KbNO$_3$, 
of which temperature 
is servo-controlled for noncritical type-I phase matching 
and is antireflection coated at 860nm and 430nm. 
The OPO cavity is locked to the degenerate frequency, $\omega_0$, 
by the Pound-Drever-Hall method\cite{PoundDrever}. 
Reflected photons of the locking beam from the surfice of the crystal 
can generate noise. 
A mechanical chopper was used to solve the problem. 
The output of the OPO and the reflected photons do not simultaneously 
go through the chopper. 
Therefore the reflected photons of the locking beam 
cannot reach detectors to produce noise.
This technique was used in Ref.\cite{OuPRL,OuPRA}. 
Dichroic mirrors and a prism are used to remove the photons of wavelength 
430nm from the signal of wavelength 860nm. 
The output from the OPO is split into two with a 50/50 beamsplitter. 
The two beams are coupled to optical fibers 
and detected with avalanche photodetectors
(APD, EG\&G SPCM-AQR-14). 
The coincidence counts of the signals from the two APDs are 
measured with a time-to-amplitude converter(TAC, ORTEC 567) and 
a multichannel analyzer(MCA, NAIG E-562).

\begin{figure*}[htbp]
	\includegraphics[width=17cm, height=7.5cm]{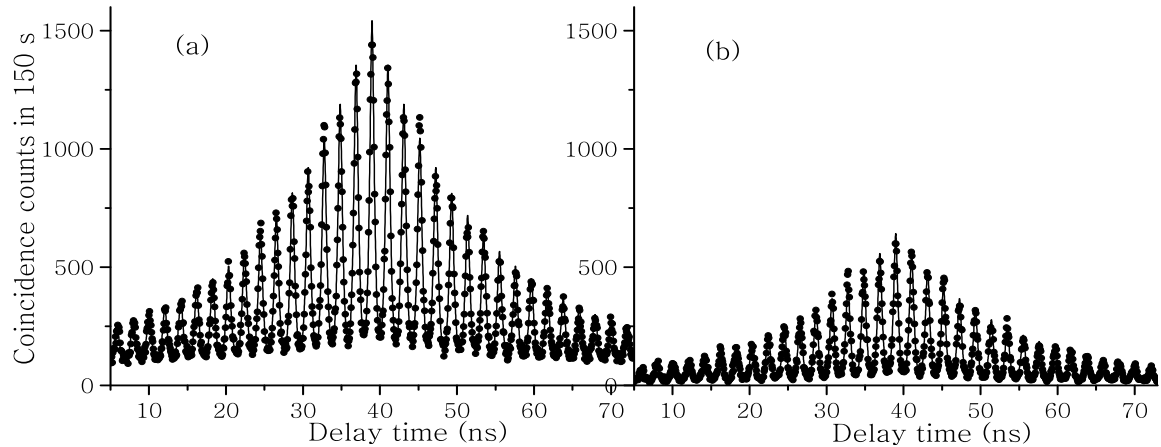}
	\caption{Experimental results of the coincidence counting of 
					the output from the OPO with a round-trip length of 560mm. 
					The pump power is about 
					(a)13$\mu$W and (b)6.5$\mu$W. The lines are 
					the fitted curves of the data to Eq.(\ref{coincidence}). 
					The fitting parameters are as follows:
					(a)$\tau_F=2.07$ns, $T_R=285$ps, $\Omega_c /(2\pi )=11$MHz, 
					$C_1=1446$, $C_2=0.067$, and $\tau_0=39$ns;
					(b)$\tau_F=2.07$ns, $T_R=274$ps, $\Omega_c /(2\pi )=11$MHz, 
					$C_1=626$, $C_2=0.025$, and $\tau_0=39$ns.
					}
	\label{result}
\end{figure*}

Figure.\ref{result} shows the experimental results. 
The circles in Figs.\ref{result} (a) and (b) represent 
the coincidence counts 
at about 13-$\mu$W and 6.5-$\mu$W pumping levels, respectively. 
The lines in Fig.\ref{result} are the fitted curves of the data 
to Eq.(\ref{coincidence}) derived as follows. 
We have assumed 
$p(\tau ) \propto \exp (- 2|\tau | \ln 2 / T_R )$ 
to obtain the best fitting of the data, where 
the resolving time of detectors, $T_R$, is 
defined as the full width at half maximum(FWHM) of $p(\tau )$. 
From Eq.(\ref{correlationav2}), 
the number of the coincidence counts, $\Gamma_c(\tau )$, is calculated as 
\begin{widetext}
\begin{align}
\Gamma_c(\tau ) 
=
C_1
\left[
C_2
+
e^{-\Omega_c |\tau -\tau_0|} 
\sum_n 
\left(
1+\frac{2 |\tau - n\tau_F -\tau_0| \ln 2}{T_R}
\right)
\exp
\left(
-
\frac{2 |\tau -n\tau_F -\tau_0| \ln 2}{T_R}
\right)
\right], 
\label{coincidence}
\end{align}
\end{widetext}
where $C_1$ and $C_2$ are constants, both of which are proportional to 
the pump power; 
$\tau_0$ denotes an electric delay. 
The results of the fitting are as follows:
(a)$\tau_F=2.07$ns, $T_R=285$ps, $\Omega_c /(2\pi )=11$MHz, 
$C_1=1446$, $C_2=0.067$, and $\tau_0=39$ns;
(b)$\tau_F=2.07$ns, $T_R=274$ps, $\Omega_c /(2\pi )=11$MHz, 
$C_1=626$, $C_2=0.025$, and $\tau_0=39$ns. 
The round-trip time of the OPO obtained by the fitting, 2.07ns, 
is nearly equal to that calculated from the cavity length, 1.9ns. 
From the bandwidth of the OPO obtained by the fitting, 11MHz, 
we can estimate the loss of the OPO is about 4\%. 
The ratios of the coefficients, $C_1$ and $C_2$, in Fig.\ref{result} (a) 
to those in Fig.\ref{result} (b) 
are 2.3 and 2.7, 
respectively. 
These are nearly equal to the ratio between 
the pump power of Figs.\ref{result} (a) and (b), 2. 
This is consistent with the above theory. 
In addition, it was confirmed that the signal was much more intense than 
unwanted noises such as reflected photons 
of the locking beam, the pump beam, and scattered photons of the laser. 
Thus we concluded that
the term of higher-order than $|\epsilon |^2$ induces 
the signal independent of the delay time, $\tau$. 
The deviations of the observed ratios of the coefficients 
are considered to be due to the fluctuation of the pump power.

In conclusion, 
we have observed the oscillatory correlation function of the 
multimode two-photon pairs produced with the OPO. 
The cavity length of the OPO is relatively long(560mm) 
because 
the observation of the oscillation requires the longer round-trip time of 
an OPO than the resolving time($\sim 300$ps) of detectors.
The fitting of the experimental data to a theoretical curve 
is excellent as seen in Fig.2. 
The multimode two-photon state from an OPO 
has the following characteristic feature. 
One of a multimode two-photon pair from an OPO 
can come relatively far from the other photon
like a single-mode two-photon pair from an OPO. 
On the other hand, 
one of the multimode two-photon pair from the OPO can only come 
at specific delay time from the other photon
like a single-pass down-converted two-photon pair. 
It is expected the multimode two-photon state 
has possibility of being applied to quantum information technology.

The authors thank Messrs. 
M. Ueki, S. Otsuka, and Y. Nanjo for their help 
to design and manufacture the experimental equipment.


\begin{thebibliography}{22}
	\bibitem{Weinberg}
						D. C. Burnham and D. L. Weinberg, 
						Phys. Rev. Lett. \textbf{25}. 84 (1970)
	\bibitem{MandelPDCtheoryA}
						C. K. Hong and L. Mandel
						Phys. Rev. A \textbf{31}, 2409 (1985)  
	\bibitem{MandelPDCtheoryL}
						S. Friberg, C. K. Hong, and L. Mandel, 
	          Phys. Rev. Lett. \textbf{54}, 2011 (1985)  
	\bibitem{Ghosh}
						R. Ghosh and L. Mandel, 
						Phys. Rev. Lett. \textbf{59}, 1903 (1987)
	\bibitem{Hong}
						C. K. Hong, Z. Y. Ou, and L. Mandel, 
						Phys. Rev. Lett. \textbf{59}, 2044 (1987)
	\bibitem{Ou1989}
						Z. Y. Ou and L. Mandel, 
						Phys. Rev. Lett. \textbf{62}, 2941 (1989)
	\bibitem{Ou1990}
						Z. Y. Ou, L. J. Wang, X. Y. Zou, and L. Mandel, 
						Phys. Rev. A \textbf{41}, 566 (1990)
	\bibitem{OuFranson}
						Z. Y. Ou, X. Y. Zou, L. J. Wang, and L. Mandel, 
						Phys. Rev. Lett. \textbf{65}, 321 (1990)
	\bibitem{Chiao}
						P. G. Kwiat, W. A. Vareka, C. K. Hong, H. Nathel, and R. Y. Chiao, 
						Phys. Rev. A \textbf{41}, 2910 (1990)
	\bibitem{Zou}
						X. Y. Zou, L. J. Wang, and L. Mandel, 
						Phys. Rev. Lett. \textbf{67}, 318 (1991)
	\bibitem{BouwmeeterTeleportation}
						D. Bouwmeester, J.-W. Pan, K. Mattle, M. Eibl, 
						H. Weinfurter, and A. Zeilinger, 
						Nature (London) \textbf{390}, 575 (1997)
	\bibitem{GhostImaging}
						D. V. Strekalov, A. V. Sergienko, D. N. Klyshko, and Y. H. Shih, 
						Phys. Rev. Lett. \textbf{74}, 3600 (1995) 
	\bibitem{ShihLithography}
						M. D'Angelo, M. V. Chekhova, and Y. Shih, 
						Phys. Rev. Lett. \textbf{87}, 013602 (2001) 
	\bibitem{Milburn}
						G. J. Milburn and D. F. Walls, 
						Opt. Commun. \textbf{39}, 401 (1981) 
	\bibitem{Collett}
						M. J. Collett and C. W. Gardiner, 
						Phys. Rev. A \textbf{30}, 1386 (1984)
	\bibitem{GardinerOC}
						C. W. Gardiner and C. M. Savage, 
						Opt. Commun. \textbf{50}, 173 (1984)
	\bibitem{KimbleSqueeze}
						L.-A. Wu, H. J. Kimble, J. L. Hall, and H. Wu, 
						Phys. Rev. Lett. \textbf{57}, 2520 (1986)
	\bibitem{OuPRL}
						Z. Y. Ou and Y. J. Lu,
						Phys. Rev. Lett. \textbf{83}, 2556 (1999) 
	\bibitem{OuPRA}
						Y. J. Lu and Z. Y. Ou, 
						Phys. Rev. A \textbf{62}, 033804 (2000) 
	\bibitem{Ou}
						Z. Y. Ou,
						Phys. Rev. A \textbf{37}, 1607 (1999) 
	\bibitem{OuStatistics}
						Y. J. Lu and Z. Y. Ou, 
						Phys. Rev. Lett. \textbf{88}, 023601 (2002)
	\bibitem{PoundDrever}
						R. W. Drever, J. L. Hall, F. V. Kowalski, J. Hough, 
						G. M. Ford, A. J. Munley, and H. Ward, 
						Appl. Phys. B \textbf{31}, 97 (1983)
\end{thebibliography}
\end{document}